\documentstyle[prd,aps]{revtex}
\begin{document}
\draft

\def\beq{\begin{equation}}
\def\eeq{\end{equation}}
\def\bea{\begin{eqnarray}}
\def\eea{\end{eqnarray}}
\def\simlt{\stackrel{<}{{}_\sim}}
\def\simgt{\stackrel{>}{{}_\sim}}
\input epsf
\twocolumn[\hsize\textwidth\columnwidth\hsize\csname
@twocolumnfalse\endcsname

\title{Baryogenesis at Low Reheating Temperatures}
\author{Sacha  Davidson$^{(1)}, $Marta Losada$^{(2)}$ and Antonio
Riotto$^{(3,4)}$}

\address{$^{(1)}${\it Theoretical Physics,  Oxford
University, 1 Keble Road, Oxford, OX1 3NP, UK }}
\address{$^{(2)}${\it Centro de Investigaciones, 
Universidad Antonio Nari\~{n}o, Cll. 57 No. 37-71, Santa Fe de Bogot\'{a},
Colombia }}
\address{$^{(3)}${\it Scuola Normale Superiore, Piazza dei Cavalieri 7, 
I-56126 Pisa, Italy}}
\address{$^{(4)}${\it INFN, Sezione di Pisa, 
I-56127 Pisa, Italy}}

\date{January, 2000}
\maketitle
\begin{abstract}
We note  that the maximum temperature 
during   reheating  
 can be much greater
than the reheating  temperature $T_r$  at which the Universe
becomes radiation dominated.
We show that the Standard Model 
anomalous $(B+L)$-violating processes
can therefore be   in thermal equilibrium for  
 1 GeV  $\simlt T_{r}\ll 100$ GeV. Electroweak
 baryogenesis could work and
the traditional upper bound on the Higgs mass 
coming from the 
requirement of the  preservation of the baryon asymmetry 
  may be  relaxed. 
 Alternatively, the baryon asymmetry may be
reprocessed by sphaleron transitions either  from a 
 $(B-L) $ asymmetry generated by the Affleck-Dine mechanism or 
from  a chiral  asymmetry between $e_R$ and $e_L$ 
 in a $B-L = 0$ Universe.
Our findings are also relevant to the production
of the baryon asymmetry in  large extra
dimension models.
\end{abstract}
\vskip2pc]

\def\simlt{\stackrel{<}{{}_\sim}}
\def\simgt{\stackrel{>}{{}_\sim}}

{\it Introduction.}~~Theories that explain the tiny difference between the 
number density of baryons and  antibaryons $ -$  about $10^{-10}$ if normalized
 to the entropy density of the Universe $-$  represent perhaps 
 the best example of the  interplay between particle physics and cosmology.
  Until now, many mechanisms for the generation of the baryon asymmetry
 have been proposed \cite{reviews}. 
Baryogenesis at the electroweak
scale has been of recent interest,
and is  attractive because it can be tested 
at  current and future accelerator experiments.
  On the other hand, 
 we know that 
the flatness and the horizon problems of the standard big bang
cosmology are elegantly solved if during the evolution of the early
Universe the energy density is dominated by some form of
vacuum energy, and comoving scales grow quasi-exponentially \cite{revinf}.
This naturally generates the observed large scale
density and temperature fluctuations.
This  inflationary stage 
can be parametrised by the evolution of some 
scalar field $\phi$,  the {\it inflaton}, 
which is initially displaced from the minimum of its potential. 
Inflation ends when the potential energy associated with the inflaton
field becomes smaller than the kinetic energy of the field.  
The  low-entropy cold Universe
dominated by the energy of coherent motion of the $\phi$ field must then be
transformed into a high-entropy hot Universe dominated by
radiation. This process 
has been dubbed
{\it reheating}. Of particular interest   is
 a quantity known as the reheating 
temperature $T_{r}$,
 defined  such that
the energy density of the Universe 
when it  becomes dominated
by radiation is $ \propto T_r^4$. 
Notice that the Universe might have gone through
further   processes 
       of reheating if -- after inflation -- the energy density
of the Universe happened to be dominated by the the coherent oscillations
of some generic  weakly-coupled
scalar fields,
{\it e.g.} some moduli fields
which are ubiquitous in string and supersymmetric theories.

A common assumption in baryogenesis models
 is that the post-inflationary Universe
contained 
a     plasma in thermal equilibrium with
initial 
temperature $T$ much larger than (or at least of
order of)
  the electroweak scale. This is required to 
 have
 acceptable initial conditions for the most popular baryogenesis
 mechanisms, and
to  take advantage of 
the Standard Model (SM) anomalous $(B+L)$-violation.

This assumption seems
so natural that it is rarely questioned.
 However,  low reheating temperature scenarios are
particularly welcome      if one wishes to avoid the overproduction of
dangerous relics  at
(pre)heating stage \cite{gra} after     inflation 
(such as gravitinos and moduli fields),
or at reheating (gravitons
in  models with large extra dimensions \cite{gia}). Apart from 
these speculative arguments,         
we  do not know the history of the 
 observable Universe before the epoch
of nucleosynthesis
---all we know 
experimentally 
is that $T_r \simgt$  
1 MeV.

 The three required ingredients for baryogenesis are
baryon number violation, C and CP violation and
out-of-equilibrium dynamics.  
It is not easy to generate 
the baryon asymmetry in  a Universe that
reheats to a low temperature 
because    the first and third ingredients 
are hard to come by \cite{bd}: it is difficult to introduce baryon
number violation at low temperatures without
contradicting laboratory bounds on $B$
violation, and the Universe is
expanding so slowly at low temperatures that
it is very close to equilibrium. 
There are nonetheless some models
for baryogenesis in cold Universes\cite{bd,cU}.

The possibility of using anomalous electroweak 
$\Delta B = \Delta L = 3$ operators
to generate the baryon asymmetry
in a low $T_r$ Universe is particularily
interesting for
Low Quantum Gravity Scale (LQGS) models \cite{gia}.
In these theories, the $(4+n)$-dimensional string
 scale $       M_s$ is well below the
$4D$
       Planck mass $M_{p}$.  Gravity is
weak on our 4-dimensional brane because it
is ``diluted'' in the $n$ compact dimensions
where ordinary matter cannot propagate.  The
usual baryogenesis mechanisms \cite{bd}
are difficult to implement in these theories
because the reheat temperature on
our brane must be low to avoid over-producing gravitons
in the large extra dimensions, and because
the laboratory bounds on baryon number violation
are significant.  If 
 every operator
not forbidden by a gauge symmetry is generated at
the quantum gravity scale with a coefficient of
order unity,   then $\Delta B = 1$     operators
capable of mediating proton
decay  need to be forbidden 
for $M_s \simlt(10^{9} - 10^{26})$ GeV
\cite{bd}. Neutron-anti-neutron oscillations
can be generated by  $\Delta B = 2$     operators,
which must be forbidden for 
 $M_s \simlt 10^{5} $ GeV.

The aim of the present Letter is to show that
baryogenesis is much less difficult than anticipated
in a Universe with a low reheating temperature (say much 
 below the electroweak scale).
Contrary to    naive expectations,   
baryogenesis scenarios using electroweak
$(B+L)$-violation  remain  
viable.
  We will show that electroweak $(B+L)$-violating processes
may be present
even though $T_r\ll 100$ GeV. This is already a surprising
result. Furthermore,   electroweak baryogenesis is possible  and 
the traditional upper bound on the Higgs mass coming from the 
requirement of the  preservation of the baryon asymmetry 
  is relaxed  because  the Universe
is expanding faster so   sphaleron  configurations  go easily 
 out of equilibrium after the electroweak phase transition (EPT)
 \footnote{see \cite{jp} for
a general phenomenological discussion of
non-standard cosmologies where the sphaleron
bound is weakened.}.
Alternatively, the anomalous $(B+L)$-violation
may reprocess an asymmetry in $(B-L)$ generated by some
other mechanism, for instance Affleck-Dine \cite{AD}.
We will also show that the electron
Yukawa coupling can be out of equilibrium
while the sphalerons are present, so a primordial
asymmetry between $e_R$ and $e_L$ in a $B-L = 0$
Universe can be transformed by the $(B  +L)  $-violation 
into a baryon asymmetry \cite{pre}.

{\it Details of the reheating stage.}~~ We now discuss
the key argument of our idea.
All our considerations are based on the
fact that 
 reheating is
far from being 
an  instantaneous process. This is a simple,
but crucial point \cite{turner,Mc}.

Suppose reheating
is due to the perturbative 
decay of a weakly-coupled scalar field $\phi$.
The latter might be the inflaton field as well 
as a modulus. The radiation-dominated phase
follows a       prolonged stage 
of coherent oscillations of $\phi$.                
During the epoch  between the initial time $H_I^{-1}$ (the time
at which the oscillations start)
and the time of reheating $\Gamma_\phi^{-1} $,  where
$\Gamma_\phi \equiv \alpha_\phi M_\phi$ 
is the decay rate of the field, the energy
density per unit comoving volume
of the scalar field $\phi$ decreases  slowly 
as $e^{-\Gamma_\phi t}$ while $\phi$
decays  into lighter states.
For low reheat temperatures,  
the decay products of the scalar field 
       thermalize rapidly      \cite{ckr,enq}.
 As the  coherent $\phi$ oscillations 
gradually
 decay, the temperature of the Universe
does not scale as $T\sim a^{-1}$ (as in the radiation-dominated era), but
follows a different law~\cite{turner,ckr} $: 
T=T_{ m  }f(a)$. Here 
\beq
T_{m  }= 0.54~
\frac{g_*^{1/8}(T_{r })}{g_*^{1/4}(T_{m  })}~(M_{p}H_I)^{1/4}~
T_{r }^{1/2} \simeq \frac{T_r}{\alpha_{\phi}^{1/4}}
\eeq
and    $
f(a)\equiv K \left( a^{-3/2}-a^{-4} \right)^{1/4}$, 
$K \equiv    1.3
({g_*(T_{m  })}/{g_*(T)})^{1/4}$.
The function $f(a)$ grows until $a_0=(8/3)^{2/5}$, where it reaches its
maximum $f(a_0)=1$, and then decreases
as $f\sim K a^{-3/8}$. Therefore, for $a>a_0$, the temperature
can be approximated by $T\simeq T_{m  }~K ~a^{-3/8}$.
This result shows that, during the phase before reheating, the temperature
reaches a maximum temperature $T_m$ and then 
has a less steep dependence on the scale factor $a$ than in the 
radiation-dominated era. 
The Hubble rate is
\beq
\label{rate}
 H\simeq \sqrt{ \frac{ 8 \pi g_*(T)}{3}} ~\frac{T^2}{ M_{p}}
\frac{g_*^{1/2}(T) T^2}{g_*^{1/2}(T_r) T_r^2},
\eeq
 and -- 
at a given temperature -- 
the expansion is faster the smaller is the reheat temperature.
Therefore $T_r$ is not 
the maximum  temperature obtained in the
universe during reheating. Note that this
should be qualitatively true  of any model
with a low $T_r$, and does not
depend on the details of reheating. 
 The maximum temperature 
can be much larger than $T_{r }$
provided that  $H_I\gg
T_{r }^2/M_{p}$; for instance $T_{m  }
\sim         10^5$ GeV for $H_I\sim$ 1 TeV and $T_{r}\sim$ 1 GeV.
This means that
anomalous $(B+L)$-violation
may be in equilibrium  even though the
reheat temperature is very low.
We also see that for temperatures
larger than $T_r$, the 
expansion rate is faster than for a
radiation-dominated Universe at a given temperature $T$.

{\it Electroweak baryogenesis.}~~ 
The fundamental idea of electroweak baryogenesis  is to produce
 asymmetries in some local charges which are (approximately) conserved
 by the interactions inside the walls of the expanding 
         bubbles formed during
the EPT.   
   Local departure from thermal 
equilibrium is attained inside the walls.
Local charges diffuse into the unbroken phase where  baryon number
 violation is active thanks to the unsuppressed $(B+L)$-violation
 \cite{krs}.  This converts the asymmetries into baryon asymmetry,
 because the state of minimum free energy is attained for nonvanishing 
baryon number. Finally, the baryon number flows into the 
broken phase where it  would be  erased
by unsuppressed  
sphaleron transitions  unless
 $\langle h(T_c)\rangle/T_c\simgt 1$, where $\langle h(T_c)\rangle$ is the
vacuum expectation value of the Higgs field at   the critical
temperature $T_c\sim$ 100 GeV \cite{boundsph}.
 Naively one expects that
the bound  $\langle h(T_c)\rangle/T_c\simgt 1$---
   obtained supposing that the electroweak
phase transition takes place in a radiation-dominated phase 
-- to translate into an upper bound on the Higgs 
mass in the SM or its extensions.
For the SM, two-loop perturbative results 
give an upper bound in the Higgs mass $m_h \simlt 45$ GeV. 
However,  nonperturbative results 
give the drastically different conclusion 
that {\it no Higgs mass} can satisfy
the above bound 
for a top mass $m_t=175$ GeV \cite{KLRS-SM}. In  the
 Minimal Supersymmetric
 Standard Model (MSSM), given the current 
LEP bound  on the Higgs mass, the so-called   
light-stop  mechanism is required to have 
sphaleron transitions out of equilibrium 
in the broken phase \cite{LSM}. Thus, 
the Higgs mass and the lightest stop mass define 
the allowed region in parameter space. However, 
we emphasize that recent analysis have shown 
that the largest allowed Higgs mass is 
obtained from zero temperature radiative 
corrections and the upper bound on the 
Higgs mass from the sphaleron 
constraint is no longer in effect
as long as one has a sufficiently 
light stop $m_{\tilde{t}} \simlt  170$ GeV \cite{CQW}.

Let us now suppose  that the reheating temperature $T_r \ll T_c$.
As we have seen 
in the previous section, the
hot thermal bath  may nonetheless
reach  temperatures
$T_m\gg T_c$.   This means that 
 the    EPT may well
 proceed      before the Universe has entered the radiation-dominated
phase when reheating is completed.
 The only difference is that the 
transition takes place 
in a matter-dominated Universe whose expansion
 rate is given by Eq.  (\ref{rate}). 
                          Electroweak baryogenesis may occur
even when $T_r\ll T_c$. This is a nontrivial result. 
     The generation of the baryon asymmetry
is mediated by sphaleron transitions in the unbroken phase, 
at a rate  $\Gamma_{s  } \simeq k \alpha_W^4 T$, where
$k \simeq 0.1 (\sim $ few $ \times \alpha_W$) \cite{sph}. They
 are in equilibrium at temperatures $T\simlt 
\left(\alpha_W^4 M_{p} T_r^2\right)^{1/3}\sim 10^4\left(T_r/1 ~
{\rm GeV}\right)^{2/3}$ GeV.

Let us now elaborate on the erasure condition.
We would like to show that the requirement that
sphalerons be
out-of-equilibrium in the broken phase 
is more easily satisfied if $T_r\ll T_c$ than
in the standard cosmology. This is a
particular case of the analysis in \cite{jp}.
 At finite temperature $T$ the rate $\Gamma_s$ per unit time
and unit volume for fluctuations between neighboring minima 
with different baryon number is
 ~\cite{CLMW} 
$\Gamma_s\sim 10^5\:T^4\left(\frac{\alpha_W}{4\pi}\right)^4
\kappa \frac{\zeta^7}{B^7}
e^{-\zeta}$, where
 $\zeta(T)=E_{s}(T)/T$, 
$E_{s}(T)=\left[2m_W(T)/\alpha_W\right]B(\lambda/g^2)$
is the sphaleron energy, 
$m_W(T)=\frac{1}{2}g\langle
  h(T)\rangle$, $B \simeq 1.9$ is a function which depends weakly 
on the gauge and the Higgs quartic couplings $g$ and $\lambda$,
  $\alpha_W=g^2/4\pi =0 .033$. Requiring $\Gamma_s/T^3 \simlt H$ at
the bubble nucleation temperature $T_b$
leads to the condition on $\zeta(T_b)$,
\begin{equation}
\label{condspha}
\zeta(T_b)\simgt 7\log\zeta(T_b)+9\log 10+\log\kappa+
2\log\left(T_r/
T_b          \right),
\end{equation}
where $H$ is given in Eq. (2).
This inequality is the standard one  \cite{reviews,sph},   with one crucial
difference:
the presence of the last term. The latter 
   tells us that, if the reheating temperature
is much smaller than $T_c$ (or
equivalently the Universe is expanding very quickly)
 sphalerons go out-of-equilibrium with ease  or they are never in equilibrium
in the broken phase!  This is one of the main results of our paper.
If we assume that $\zeta(T_b) \simeq 1.2 \zeta(T_c)$
\cite{KLRS-SM},  then for $\kappa=10^{-1}$ and $T_r\sim
1(10)
$ GeV, we obtain that $\zeta(T_c)\simgt 28(33)$, which translates into
\beq
\frac{\langle 
   h(T_c)\rangle}{
 T_c}
\simgt 0.77(0.92).
\label{bound}
\eeq
  This bound 
 has to be compared
to the standard result $\langle h(T_c)\rangle/T_c\simgt 1$ obtained
for the same value of $\kappa$. This finding
clearly enlarges
the available region in parameter space where the 
sphaleron bound is satisfied
and relaxes the       
upper bound  on the stop mass in the MSSM and on the Higgs mass in other
extensions of the SM. The implication for the  SM  is that although
current LEP bounds  on the Higgs mass still rule 
out electroweak baryogenesis,  for
small values of the Higgs mass the phase transition is now strong enough for
sphaleron transitions to be suppressed. From the lattice results of Ref.
 \cite{KLRS-SM} we can determine that Eq. (\ref{bound})  implies 
that the EPT would be strong enough 
for baryogenesis for $m_h \simlt 50$ GeV. More interesting, for the MSSM 
in the region of allowed Higgs masses the new 
bound  of Eqn (\ref{bound}) 
could increase the upper bound on the stop mass by about 10 GeV
 to $m_{\tilde{t}} \simlt 180$ GeV for
all other parameters fixed. 
These and other issues are now under investigation
\cite{inprep}.

One should not claim victory too soon, though. While preserving a baryon
asymmetry is easier if $T_r\ll T_c$, the continous decays of the
scalar field $\phi$ dump  entropy into the thermal soup 
from $T_c$ to $T_r$.
Indicating by $B_c$ the baryon asymmetry to entropy density
ratio $n_B/s$ 
generated at the EPT, 
one finds that the final baryon
asymmetry is
\cite{Mc}
\beq
\frac{n_B}{s}\sim B_c\left(\frac{T_r}{T_c}
\right)^5.
\label{entropy}
\eeq
This means that, for $T_r\sim 10$ GeV,
the mechanism of baryogenesis
at the electroweak scale has to be more efficient by a factor
$\sim 10^5$ than in the standard case. This is certainly challenging, but
not impossible to achieve.  Parametrizing 
$B_c\sim \kappa \alpha_W^4\delta_{CP}f(v_w)$, one would need 
 the CP-violating phases $\delta_{CP}$
and the velocity of the bubble walls $v_w$ to be of order of 
                    unity \cite{inprep}.

{\it Reprocessing a pre-existing asymmetry.}~~
An alternative to electroweak baryogenesis when $T_r$ is low   
 is to 
 make use of the 
    anomalous electroweak $(B+L)$-violation to  transform
a pre-existing asymmetry in $(B-L)_L   $ into a baryon asymmetry \cite{FY}.
The Affleck-Dine mechanism \cite{AD} is particularly
attractive in our framework
             since it 
 can naturally generate
 a  lepton
asymmetry 
  when             the          slepton fields along
the flat directions                           relax to their
minima \cite{CDO}. This 
happens when 
the Universe
is still dominated by the $\phi$-oscillations and  
     the hot plasma
     is still at temperatures much larger than 
$T_r$.
The initial lepton
asymmetry can naturally be of order unity,    it gets
reprocessed  into baryon asymmetry by sphaleron interactions and 
is            subsequently reduced to the observed value
by the large      entropy production \cite{inprep,prob}. 

A further and 
 new possibility                     is
that the    sphalerons   can reprocess a
pre-existing asymmetry between the
$e_L$ and $e_R$ into a baryon asymmetry \cite{pre}.
This is interesting because the only
 $B$ or $L$ violation  required is
the SM   sphalerons,   but
the out-of equilibrium and CP violation
required to generate an asymmetry can
take place somewhere other than at
the EPT.                          
The idea is that the Universe
starts with $B = L = 0$, and                    an
excess of  ${e}_R$ over  anti-$e_R$
is created during the $\phi$-oscillations.
The Universe is electrically neutral, so there must be asymmetries
among other charged particles to compensate
the $e_R$ charge density. The electron Yukawa
is small, so the $e_R$ remain
out of chemical equilibrium until
late times. 
The  anomalous SM             $(B+L)$-violation
 is rapid, and acts only on left-handed
particles, among which there is
a lepton number deficit. This asymmetry
in $L_L$  will therefore be partially
transformed into a baryon asymmetry.
If the $(B+L)$-violating processes go
out of equilibrium before the $e_R$ comes
into chemical equilibrium, then this
baryon asymmetry will be preserved. 
In the standard cosmology, this is
not the case: the sphalerons go out
of equilibrium at or just after
the electroweak phase transition,
and the electron Yukawa comes into
equilibrium before this at temperatures
$\sim (
10-100)
$ TeV \cite{pre,cko}. 
However, in our scenario, the
expansion rate of the Universe
is faster, so
it could be possible to reprocess
an initial chiral asymmetry between
$e_L$ and $e_R$ into a baryon asymmetry.
We need to check that 
the $e_R$ are {\it out}
of chemical equilibrium while the   sphalerons
are in equilibrium. As previously
discussed, there will be $(B+L)$-violation
 in equilibrium above the electroweak
phase transition if $T_r \simgt
10  $ MeV. We can
estimate the rate associated with the
electron Yukawa coupling $h_e$ to be
$\Gamma_{h_e} \simeq 10^{-2} h_e^2 T$,
in which case $\Gamma_{h_e} \simgt
  H$  at
$T\simlt
  30 (T_r/{\rm GeV})^{2/3} $ GeV.
 So for $T_r\simlt$  a few GeV,
 we find that the $e_R$ do not come
into equilibrium until after the
sphalerons are out of equilibrium.
This estimate suggests that 
 an initial chiral asymmetry
between $e_R$ and $e_L$ in
a $B= L = 0$ Universe can be
reprocessed into a baryon asymmetry.
However, the $e_R$
may also be brought into chemical equilibrium
by anomalous processes, which we will
discuss in a subsequent publication \cite{inprep}.
Note that for
this mechanism, the only $B$ or $L$ violation required
is that already present in the Standard Model,
but large amounts of CP violation
or departure from equilibrium are
not required at the  EPT.

                  In conclusion, we have shown that the  simple
observation that --
  in a 
Universe with a low reheat temperature $T_r$ --
                                        the maximum temperature of
the thermal bath                            can
be much larger than $T_r$ has rich implications
for baryogenesis.
This          is extremely 
     encouraging 
 because, after all, 
observationally we   only 
 know that $T_r$ has to be larger than a few MeV to allow primordial
nucleosynthesis. 

\acknowledgements
We would like to thank Ian Kogan, Misha Shaposhnikov and
Carlos Wagner for useful conversations.

\end{document}